\documentclass[twocolumn,aps,pra,10pt]{revtex4-1}
\usepackage{bm}
\usepackage{amsfonts}
\usepackage{amssymb}
\usepackage{amsmath}
\usepackage{graphicx}
\begin{document}

\title{Comment on ``Nondispersive analytical solutions to the Dirac equation''}
\author{Iwo Bialynicki-Birula}\email{birula@cft.edu.pl}
\affiliation{Center for Theoretical Physics, Polish Academy of Sciences\\
Aleja Lotnik\'ow 32/46, 02-668 Warsaw, Poland}
\author{Zofia Bialynicka-Birula}
\affiliation{Institute of Physics, Polish Academy of Sciences\\
Aleja Lotnik\'ow 32/46, 02-668 Warsaw, Poland}

\begin{abstract}
In our Comment we question the validity of the claim made by Campos and Cabrera [Phys. Rev. Research {\bf 2}, 013051 (2020)] that their solutions of the Dirac equation in an external {\em time-dependent} electromagnetic field describe beams of electrons. In every time-dependent field, no matter how weak, which has an {\em infinite} time duration, there is continuous electron-positron pair creation and annihilation. Without the proper accounting for these processes, the mathematical solutions of the Dirac equation are not directly applicable to realistic physical situations. In particular, the time evolution of the average values $\langle x\rangle$ and $\langle y\rangle$ does not describe the electron trajectory but the motion of some combination of the electron and positron charge distributions with pathological properties ({\em zitterbewegung}).
\end{abstract}
\maketitle
In their work \cite{cc} the authors present new analytical solutions of the Dirac equation hoping that ``The analytical solutions of the Dirac equation given here provide important insights into the relativistic dynamics of electrons.'' We do not question the validity of their formulas and we praise the authors for inventing a new method to generate solutions of the Dirac equation in an electromagnetic field. However, in our opinion, the physical interpretation of the results is incomplete and this also applies to the previous work by the same group of authors \cite{rdi}.

The Dirac equation in the presence of any time-dependent electromagnetic field is not an equation describing just the evolution of a wave packet of an electron. In order to use a solution of the Dirac equation to describe the motion of the {\em electron wave packets} one must be sure that the part describing the motion of electrons has been separated from the full solution. The authors of Ref.\cite{cc} have not done this. Their expectations that they describe only the relativistic dynamics of electrons are countered by a more careful analysis of the solutions of the Dirac equation in a time-dependent electromagnetic field. These solutions contain as a rule also a part describing positrons. It must be so because the same equation describes also the states of positrons. Afer all, the Dirac theory is invariant under the charge conjugation, ${\cal A}_\mu\to -{\cal A}_\mu, \psi\to\psi_{\rm C}=i\gamma^2\psi^*$, which interchanges particles and antiparticles.

We will show in what follows that even the simplest Wolkow solution of the Dirac equation \cite{wolk} has a part that describes positrons. The solutions presented in Ref.\cite{cc} are elaborate generalizations of the Wolkow solution and they have the same property.

The simplest electromagnetic wave, the plane, mono\-chromatic linearly polarized wave is described by the vector potential $A_x=A\cos[\omega(t-z)] (c=1)$. The solution $\psi_{\rm W}$ of the Dirac equation in this case (for the spin pointing along the $z$ axis) is \cite{wolk}
\widetext
\begin{align}\label{wolk}
\psi_{\rm W}(x)&=e^{-ip\cdot x}\exp\left(\!-i\frac{A^2(t-z)}{4(E_p-p_z)}\!\right)
\exp\left(\!i\frac{8Ap_x\sin(\omega(t-z))-A^2\sin(2\omega(t-z))}
{8\omega(E_p-p_z)}\!\right)(u_1(p)-A\cos(\omega(t-z))u_2(p)),
\nonumber\\
u_1(p)&=\frac{1}{E_p+m}\left[\begin{array}{c}E_p+m\\0\\
p_z\\p_x+ip_y\end{array}\right],\quad u_2(p)=\frac{1}{2(E_p+m)(E_p-p_z)}\left[\begin{array}{c}p_x+ip_y\\
-(E_p+m-p_z)\\p_x+ip_y\\E_p+m-p_z
\end{array}\right].
\end{align}
\endwidetext
This solution can be expanded into an infinite sum of plane waves,
\begin{align}\label{wolk1}
\psi_{\rm W}(x)=\sum_{n=-\infty}^{n=\infty}c_ne^{-iq_n\cdot x}v_n,
\end{align}
where $q_n=\{E_p+n\omega,p_x,p_y,p_z+n\omega\}$.
Clearly, for sufficiently large and negative $n$ the sign of the energy component of the momentum four-vector $q_n$ will be negative and the corresponding wave describes a state of positron. Note that the presence of this effect does not depend on the amplitude $A$ but only on the frequency. Of course, the amplitude of the {\em n}th positron wave is proportional to $A^n$ and for small $A$ it is small. The Wolkow wave function $\psi_{\rm W}(x)$ can also be written for positrons. In this case the sign of the momentum four-vector $p$ is reversed and we would then have an admixture of electronic states.

We can put our analysis of the Wolkow solution into a broader perspective by connecting it with the general treatment of electromagnetic processes in quantum electrodynamics. The proper treatment of such processes has been discovered already by Feynman \cite{rf} who wrote the following: ``It is possible to write down a complete solution of the problem in terms of boundary conditions on the wave function, and this solution contains automatically all the possibilities of virtual (and real) pair formation and annihilation together with the ordinary scattering processes''.

\begin{figure}
\includegraphics[width=0.25\textwidth]{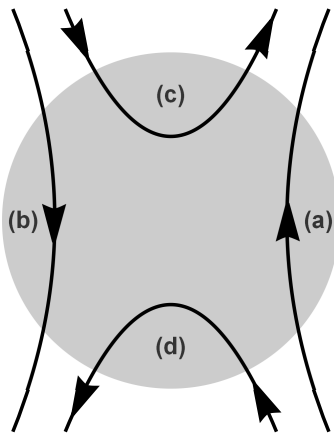}
\caption{Four different processes described by a single solution of the Dirac equation: a) Electron scattering, b) positron scattering, c) pair creation, d) pair annihilation. The shaded area represents the presence of the external electromagnetic field which is responsible for all four processes.}\label{Fig1}
\end{figure}

Following closely the Feynman approach (cf., for example, Ref.\cite{bb}), we replace the differential Dirac equation
\begin{align}\label{dir}
(i\gamma^\mu\partial_\mu-m)\psi(x)=e\gamma^\mu{\cal A}_\mu(x)\psi(x)
\end{align}
with the integral equation,
\begin{align}\label{int}
\psi(x)=\psi_{\rm F}(x)+\int\!d^4x'S_{\rm F}(x-x')\,e\gamma^\mu{\cal A}_\mu(x')\psi(x'),
\end{align}
where $\psi_{\rm F}(x)$ is a solution of the free Dirac equation and $S_{\rm F}(x-x')$ is the Feynman propagator. The integral equation guarantees the full control of the boundary conditions in the past and in the future.

It may seem at first that this procedure cannot be applied to the solutions studied in Ref.\cite{cc} because it presupposes that the electromagnetic field occupies only a bounded region of spacetime. However, we will show in what follows how to clear this hurdle.

The key point of Feynman's method is that {\em in empty space} the separation of the solution $\psi_{\rm F}$ of the Dirac equation into the electron part (positive frequencies) and the positron part (negative frequencies) is obvious. This follows from the expansion of the Dirac bispinor into plane waves, (cf., for example Refs.\cite{bd,bb}),
\begin{align}\label{pw}
&\psi_{\rm F}(x)=\int\!\frac{d^3p}{(2\pi)^{3/2}}\sum_{s=\pm}\nonumber\\
&\times\left(u(\bm p,s)f_+(\bm p,s)e^{-ip\cdot x}+v(\bm p,s)f_-(\bm p,s)e^{ip\cdot x}\right),
\end{align}
where $p\cdot x=E_pt-{\bm p}\!\cdot\!{\bm r}$ and $E_p=\sqrt{m^2+{\bm p}^2}$. An expansion into positive and negative frequencies holds also for the Feynman propagator,
\begin{align}\label{fpr}
&S_{\rm F}(x\!-\!x')=\theta(t\!-\!t')S^{(+)}(x\!-\!x')+\theta(t'\!-\!t)S^{(-)}(x\!-\!x'),\nonumber\\
&S^{(\pm)}(x\!-\!x')=\pm i(i\gamma^\mu\partial_\mu+m)\int\!\frac{d^3p}{2(2\pi)^3E_p} e^{\mp ip\cdot(x\!-\!x')}.
\end{align}
Thus, the solution $\psi(x)$ of the integral equation (\ref{int}) can be split into four parts,
\begin{align}\label{split}
\psi(x)=\frac{1}{2}\left(\psi_a(x)+\psi_b(x)+\psi_c(x)+\psi_d(x)\right),
\end{align}
where
\begin{subequations}
\begin{align}\label{abcd}
&\psi_a(x)=\psi_{\rm F}^{(+)}(x)\nonumber\\
&+\int\!d^4x'\theta(t-t')S^{(+)}(x-x')\,e\gamma^\mu{\cal A}_\mu(x')\psi(x'),\\
&\psi_b(x)=\psi_{\rm F}^{(-)}(x)\nonumber\\
&+\int\!d^4x'\theta(t'-t)S^{(-)}(x-x')\,e\gamma^\mu{\cal A}_\mu(x')\psi(x'),\\
&\psi_c(x)=\psi_{\rm F}^{(-)}(x)\nonumber\\
&+\int\!d^4x'\theta(t'-t)S^{(+)}(x-x')\,e\gamma^\mu{\cal A}_\mu(x')\psi(x'),\\
&\psi_d(x)=\psi_{\rm F}^{(+)}(x)\nonumber\\
&+\int\!d^4x'\theta(t-t')S^{(-)}(x-x')\,e\gamma^\mu{\cal A}_\mu(x')\psi(x').
\end{align}
\end{subequations}
Each part in Eqs.(6) describes one of the four distinct processes depicted in Fig.1.
\begin{figure}[b]
\includegraphics[width=0.48\textwidth,height=0.12\textheight]{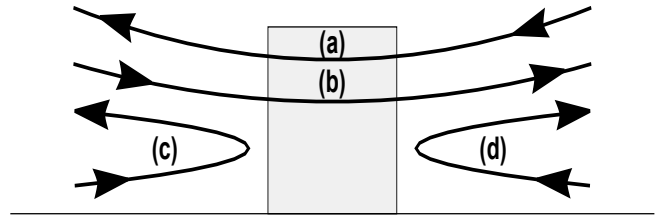}
\caption{Four different processes described by a single solution of the one-dimensional Schr\"odinger equation: a) Transmission right to left, b) transmission left to right, c) reflection from the left, d) reflection from the right. The potential is responsible for all four processes.}\label{Fig2}
\end{figure}
From the conceptual point of view this problem has a close resemblance to the quantum-mechanical problem of scattering by a potential barrier in one dimension (cf., for example, Refs.\cite{ls,em}). Also in this case a general  solution of the Schr\"odinger equation describes all four processes of transmission and reflection of the probability wave shown in Fig. 2.

The separation of the solution of the Dirac equation into four parts describing different processes when the field occupies all spacetime can be accomplished with the use of the relativistic extension of the time-of-flight method. In the nonrelativistic case this method can be used to separate the momentum components in the Schr\"odinger wave function as was vividly described by Feynman and Hibbs \cite{fh}. The time-of-flight method has also practical applications as a diagnostic tool. It was used, for example, in Nobel-prize-winning experiments \cite{cw,k}. The state of the condensate was determined by opening the trap and obtaining the information about the momentum distribution from the free evolution of the system.

The adaptation of the time-of-flight method to the study of wave functions obeying the Dirac equation works also when the electromagnetic field extends throughout the whole spacetime, as is the case in the Wolkow solution. We take the solution $\psi({\bm r},t)$ at a certain time $t_1$ and to mimic the opening of the trap we evolve this wave function {\em without the electromagnetic field} forward in time. Similarly, we do the same in the past, evolving freely the wave function backwars in time from some time $t_2$. These freely evolving bispinors in the future and in the past will be denoted by $\psi_>({\bm r},t)$ and by $\psi_<({\bm r},t)$, respectively. The formulas describing these freely evolving bispinors are:
\begin{subequations}
\begin{align}\label{free}
\psi_>({\bm r},t)=\psi_>^{(+)}({\bm r},t)+\psi_>^{(-)}({\bm r},t), \quad t>t_1\\
\psi_<({\bm r},t)=\psi_<^{(+)}({\bm r},t)+\psi_<^{(-)}({\bm r},t), \quad t<t_2,
\end{align}
\end{subequations}
where
\begin{subequations}
\begin{align}\label{free1}
\psi_{>}^{(\pm)}({\bm r},t)=\int\!d^3r'S^{(\pm)}({\bm r-\bm r'},t-t_0)\psi({\bm r'},t_0),\quad t>t_0,\\
\psi_{<}^{(\pm)}({\bm r},t)=\int\!d^3r'S^{(\pm)}({\bm r-\bm r'},t-t_0)\psi({\bm r'},t_0),,\quad t<t_0.
\end{align}
\end{subequations}
We can unambiguously identify in this way the different components of the wave function.

The solution of the Dirac equation describes solely the electrons if the only nonvanishing component in the formula (\ref{split}) is $\psi_a(x)$. In the case of the Wolkow solution, this procedure confirms our previous finding that the function $\psi_{\rm W}$ contains also the part describing positrons because the projection of the bispinor $u_2(p)$ in (\ref{wolk}) on the bispinor $v(\bm p,s)$ that describes antiparticles in the Fourier expansion (\ref{pw}) does not vanish.

In summary, the solutions of the Dirac equation in an electromagnetic field found in Ref.\cite{cc} cannot be directly applied to the description of electron beams because these solutions describe all four different physical processes: electron scattering, positron scattering, pair annihilation, and pair creation. In order to describe only the motion of electrons one must isolate the proper part of the Dirac bispinor. This can be done, for example, in the spirit of the Feynman analysis of scattering problems in the external, time-dependent electromagnetic field. The physical interpretation of the solutions according to the Feynman interpretation requires the proper identification of the four components in (\ref{split}). To this end, the solutions of the Dirac equation must be subjected to free evolutions in the future and in the past to separate positive frequency parts from the negative frequency parts. We do not exclude the possibility that there exist also a physical interpretation of the solutions obtained in Ref.\cite{cc} which does not require such a separation. We only question the validity of the claims made by the authors that their solutions describe just the motion of electrons. These solutions describe complicated physical processes involving not only electrons but also positrons.


\begin{thebibliography}{99}
\bibitem{cc} A. G. Campos and R. Cabrera, Nondispersive analytical solutions to the Dirac equation, Phys. Rev. Res. {\bf 2}, 013051 (2020).
\bibitem{rdi} A. Campos, R. Cabrera, H. A. Rabitz, and D. I. Bondar, Analytic Solutions to Coherent Control of the Dirac Equation, Phys. Rev. Lett. {\bf 119}, 173203 (2017).
\bibitem{wolk} D. M. Wolkow, \"Uber eine Klasse von L\"osungen der Diracschen Gleichung, Z. Phys. {\bf 94}, 250 (1935).
\bibitem{rf} R. P. Feynman, The theory of positrons, Phys. Rev. {\bf 76}, 749 (1950).
\bibitem{bb} I. Bialynicki-Birula and Z. Bialynicka-Birula, {\em Quantum Electrodynamics}, (Pergamon, London, 1975; Elsevier, Amsterdam, 2013); http://www.cft.edu.pl/\~{}birula.
\bibitem{bd} J. D. Bjorken and S. D. Drell, {\em Relativistic Quantum Mechanics}, (McGraw-Hill, New York, 1965).
\bibitem{ls} L. I. Schiff, {\em Quantum Mechanics}, (McGraw-Hill, New York, 1968), p. 100.
\bibitem{em} E. Merzbacher, {\em Quantum Mechanics}, (Wiley, New York, 1970).
\bibitem{fh} R. P. Feynman and Albert R. Hibbs, {\em Quantum Mechanics and Path Integrals}, (McGraw-Hill, New York, 1965).
\bibitem{cw} E. A. Cornell and C. E. Wieman, Nobel Lecture: Bose-Einstein condensation in a dilute gas, the first 70 years and some recent experiments, Rev. Mod. Phys. {\bf 74}, 875 (2002).
\bibitem{k} W. Ketterle, Nobel lecture: When atoms behave as waves: Bose-Einstein condensation and the atom laser, Rev. Mod. Phys. {\bf 74}, 1131 (2002).
\end{thebibliography}
\end{document}